\documentclass[twocolumn,showpacs,preprintnumbers,color]{revtex4}
\usepackage{graphicx}% Include figure files
\usepackage{dcolumn}% Align table columns on decimal point
\usepackage{latexsym}
\usepackage{color}
\def\der{\partial}
\def\vf{\varphi}
\def\dvf{\delta\vf}

\def\a{\alpha}
\def\b{\beta}
\def\by{y_+}

\def\e{{\mathrm {e}}}

\begin{document}
%\tighten
%\draft
%\renewcommand{\topfraction}{0.8}

\input epsf

\preprint{LAPTH-998/03, CERN-TH/2003-226}

\title{Goldberger-Wise variations: stabilizing brane models with a bulk scalar}

\author{Julien Lesgourgues$^{1,2}$ and Lorenzo Sorbo$^1$}

\affiliation{$^1$Laboratoire de Physique Th\'eorique LAPTH, F-74941
Annecy-le-Vieux Cedex, France\\
$^2$TH-Division CERN, CH-1211 G\`eneve 23, Switzerland}

\date{26 January 2004}

\begin{abstract} 
Braneworld scenarios with compact extra--dimensions need the volume of
the extra space to be stabilized. Goldberger and Wise have
introduced a simple mechanism, based on the presence of a bulk scalar
field, able to stabilize the radius of the Randall--Sundrum model.
Here, we transpose the same mechanism to generic single--brane and
two--brane models, with one extra dimension and arbitrary 
scalar potentials in the bulk and on the branes. 
The single--brane construction turns out to be
always unstable, independently of the bulk and brane
potentials. In the case of two branes, we derive some generic criteria
ensuring the stabilization or destabilization of the system.
\end{abstract}


\maketitle

%%%%%%%%%%%%%% 
%Introduzione%
%%%%%%%%%%%%%%

Models with compact extra--dimensions need the volume of the extra space to be stabilized. This is a typical problem of string--theoretical constructions, in which the existence of a mass term for moduli is vital for the viability of the models. The inevitability of $9$ or $10$ spatial dimensions in string theory provides one of the main motivations for the recent interest about extra--dimensions, although many scenarios that are not necessarily linked to string theory have also been envisaged. Among these, the one proposed by Randall and Sundrum~\cite{RS} has been object of intense study. One of the (many) features that make the scenario so interesting is the presence of a single extra--dimension, that allows for a detailed -- and not too complicated -- analysis of the model and of its possible modifications.

Soon after the proposal of~\cite{RS} (that was not addressing the
problem of the stabilization of the extra--space), Goldberger and
Wise~\cite{GW} have considered a simple mechanism, based on the
presence of a bulk scalar field, able to stabilize the radius of the
Randall--Sundrum model, as was proved in detail in~\cite{TM}
and~\cite{CGK}. Since then, the literature has been enriched by many
works about brane models in presence of bulk scalars.
The importance of radion stabilization is related to the phenomenological
consistency of the models \cite{pheno} as well as to possible
observational consequences \cite{trans}. For some specific form of
the scalar field potential the behavior of the system could be
reconstructed~\cite{solu}, by solving the bulk Einstein equations. In
most cases, however, such solution can be obtained only either in the
limit in which the backreaction of the scalar field on the background
geometry is negligible, or numerically. For instance, the dynamics
which leads various examples of two--brane models to stable
equilibrium was recently studied numerically in ~\cite{marco}.

In the present paper we will analyze the spectrum of scalar
perturbations for $(4+1)$--dimensional brane models in the presence of
a bulk scalar field. Our main goal is to provide criteria that can
tell us in which case tachyon modes emerge: a system with tachyon
modes is unstable and would quickly decay into some different ground
state. We analyze the equations describing scalar perturbations around
a background with $(3+1)$--dimensional flat branes, without referring
to any specific potential for the scalar field, and without making any
approximation. We classify models only on the basis of the behavior of
the background metric and scalar field.  
We find that,
  under general assumptions, models with only one brane always feature
  tachyon modes, consistently with the conjecture in~\cite{TM}. 
In the case of models that include two branes, we provide general
criteria about the presence or the absence of tachyon modes in the
spectrum of the theory.

%%%%%%%%%%%%%%%%%%%%%%%%%%%%%%%%%%%
\section{Setup and main equations}%
%%%%%%%%%%%%%%%%%%%%%%%%%%%%%%%%%%%

We consider a five--dimensional spacetime, with four--dimensional
Minkowski sections (spanned by the coordinates $x^\mu$) and one
compact extra--dimension $y$. Two three--branes are located at the
fixed points $y=y_-$ and $y=y_+>y_-$ of a $Z_2$ symmetry
$y\leftrightarrow 2\,y-y_+$, $y\leftrightarrow 2\,y_--y$. In section
II we will specialize to the case in which only one brane is
present. The matter content of the model is given by a bulk scalar
field $\vf$, with a bulk potential $V\left(\vf\right)$ and additional
potential terms $U_\pm\left(\vf\right)$ localized on the branes. 
A possible cosmological constant in the bulk and eventual
brane tensions
are included in the definition of $V$ and $U_{\pm}$.
The (background) metric for such system can be written as 
\begin{equation}\label{backmet}
ds^2=\bar{g}_{MN}\,dx^M dx^N\equiv a\left(y\right)^2\,\left(\eta_{\a\b}\,dx^\a dx^\b+dy^2\right)\,\,,
\end{equation}
with $\eta_{\a\b}={\rm {diag}}(-1,\,1,\,1,\,1)$.

Denoting by a prime the derivative with respect to the extra
coordinate $y$, and defining $H\left(y\right)\equiv
a'\left(y\right)/a\left(y\right)$, we write the background
equations of motion as
\begin{eqnarray}\label{backdyn}
&&\vf_0''+3\,H\,\vf_0'-a^2\,\frac{dV}{d\vf}\left(\vf_0\right)=
\, \delta(y_{\pm}) \,\,  a \, \frac{d U_{\pm}}{d \vf} (\vf_0)
\,,\nonumber\\
&&H'=H^2-\frac{\kappa^2}{3}\,\vf_0'{}^2
\, - \, \delta(y_{\pm}) \, \frac{a}{3} \, \kappa^2 
U_{\pm} (\vf_0)\,,
\end{eqnarray}
with the constraint 
\begin{equation}\label{backcons}
H^2=\frac{\kappa^2}{6}\,\left[\frac{1}{2}\,\vf_0'{}^2-a^2\,V\left(\vf_0\right)\right] \,.
\end{equation}
$\kappa^2$ represents the five--dimensional Newton constant.

The Israel junction conditions, together with the integration of the Klein-Gordon equation across the branes, give the boundary conditions at the brane locations 
\begin{eqnarray}\label{backisr}
H\left(y_{\pm}\right)=\pm\frac{\kappa^2}{6}\,a\left(y_\pm\right)\,U_\pm\left(\vf_0\left(y_\pm\right)\right)\,\,,\nonumber\\ 
\vf_0'\left(y_{\pm}\right)=\mp\frac{1}{2}\,a\left(y_\pm\right)\,\frac{dU_\pm}{d\vf}\left(\vf_0\left(y_\pm\right)\right)\,\,.
\end{eqnarray}

In order to study the stability of this system, we perturb the metric and the scalar field as $g_{MN}=\bar{g}_{MN}+h_{MN}$, $\vf=\vf_0+\dvf$, with
\begin{eqnarray}
h_{\a\b}&=&2\,a\left(y\right)^2\,\left(\psi\,\eta_{\a\b}-\der_\a\der_\b E \right)\nonumber\\
h_{\a5}&=&h_{5\a}=-a\left(y\right)^2\,\der_\a B\nonumber\\
h_{55}&=&2\,a\left(y\right)^2\,\phi\,\,.
\end{eqnarray}
Since we expect possible instabilities to come from $(3+1)$--dimensional scalar modes (the radion), we limit our analysis to scalar perturbations.

The system  turns out to be fully described by the following three gauge invariant quantities
\begin{eqnarray}
\Psi&\equiv& \psi+H\,\left(B-E'\right)\,,\nonumber\\
\Phi&\equiv& \phi+H\,\left(B-E'\right)+\left(B-E'\right)'\,,\nonumber\\
\Delta&\equiv& \dvf+\vf_0'\,\left(B-E'\right)\,,
\end{eqnarray}
and the linearized bulk Einstein equations reduce, in terms of these gauge invariants, to the constraints
\begin{eqnarray}
\label{const1}
\Phi&=&-2\,\Psi\,,\\
\label{const2}
-\vf_0'\,\Delta&=&3\,\left(\Psi'+2\,H\,\Psi\right)\,\,,
\end{eqnarray}
and to the dynamical equation (written here only in the bulk,
we will treat the boundaries separately)
\begin{equation}\label{dyn} 
\Box \Psi+4H\Psi'+\left(6H^2\!\!+2H'\right)\!\Psi=
\frac{1}{3}[\vf_0'\Delta'\!
-a^2\frac{dV}{d\vf}\!\left(\vf_0\right)\Delta],
\end{equation}
from which we derive a wave equation for $\Psi$ in the bulk
\begin{eqnarray}\label{eqpert}
\Psi'' \!+\Box\Psi+\left(3H-2\,\frac{\vf_0''}{\vf_0'}\right)\Psi'\!
+4\left(H'\!-H\frac{\vf_0''}{\vf_0'}\right)\Psi=0,\nonumber\\
\end{eqnarray}
where the box denotes the four--dimensional D'Alembert operator.

The last equation can be rewritten as
\begin{equation}\label{equ}
u''+\Box\,u-\frac{\theta''}{\theta}\,u=0\,,\qquad \theta\equiv \frac{H}{a^{3/2}\,\vf_0'}\,\,.
\end{equation}
where we defined, in analogy with the usual four--dimensional cosmological case~\cite{MFB}, $u\equiv \left(a^{3/2}/\vf_0'\right)\,\Psi$.

In general, there can be at least one point $y_0$ in which
$\vf_0'\left(y_0\right)=0$ and the equation of motion for $\Psi$ and
$u$ are singular. 
However, it is possible to go across the singularity using
a third quantity whose equations of motion remain regular
\cite{finelli}.  This is the
equivalent (for our $5$-dimensional setup) of the Mukhanov
variable~\cite{mukhanov}, i.e.
\begin{equation}\label{defv}
v\equiv a^{3/2}\,\left(\Delta-\frac{\vf_0'}{H}\,\Psi\right)=-3\,\theta\,\left(\frac{u}{\theta}\right)'\,\,,
\end{equation}
that obeys 
\begin{equation}\label{eqv}
v''+\Box\,v-\frac{z''}{z}\,v=0\,,\qquad z\equiv 1/\theta\,,
\end{equation}
where the ratio
\begin{eqnarray}\label{zssz}
\frac{z''}{z}&=&\frac{15}{4}\,H^2-\frac{19\,\kappa^2}{6}\,\vf_0'{}^2+a^2\,\frac{d^2V}{d\,\vf^2}+\frac{2\,\kappa^4}{9}\,\frac{\vf_0'{}^4}{H^2}+\nonumber\\
&&+\frac{4\,\kappa^2}{3}\,\frac{\vf_0'}{H}\,a^2\,\frac{d\,V}{d\,\vf}\,\,,
\end{eqnarray}
is always regular as long as $H\neq 0$. 
Going back to the metric perturbation is straightforward 
using the relation
\begin{equation}\label{vtopsi}
\Box\,\Psi = \frac{\vf_0'}{a^{3/2}} \frac{(\theta \, v)'}{3 \, \theta}\,.
\end{equation}

The boundary conditions for $\Psi$ can be obtained by perturbing the Israel
junction conditions and the Klein--Gordon equation. Besides the
perturbed quantities defined above, one has in this case also to take
into account the perturbation $\zeta_\pm$ of the position of the
branes. Defining the gauge invariant $Z_\pm\equiv
\zeta_\pm-\left[B\left(y_\pm\right)-E'\left(y_\pm\right)\right]$, the
junction conditions yield $Z_\pm=0$, whereas the Klein--Gordon
equation gives the following boundary conditions
\begin{equation}\label{bounpert}
g_\pm\,\left[\Psi'\left(y_\pm\right)+2\,H\left(y_\pm\right)\,\Psi\left(y_\pm\right)\right]+\Box \Psi\left(y_\pm\right)=0\,\,,
\end{equation}
where we have defined 
\begin{equation}\label{defgpm}
g_\pm\equiv H\left(y_\pm\right)-\frac{\vf_0''\left(y_\pm\right)}{\vf_0'\left(y_\pm\right)} \mp \frac{1}{2}\,a\left(y_\pm\right)\,\frac{d^2U_\pm}{d\,\vf^2}\left(\vf_0\left(y_\pm\right)\right)\,\,.
\end{equation}

Fourier--transformation of the above equations along the four ordinary
dimensions amounts to the substitution of the $\Box$ operator with the
mass eigenvalue $m^2$. Solving eq.~(\ref{eqpert}) with the boundary
conditions (\ref{bounpert}) leads to a discrete mass spectrum,
corresponding to the various Kaluza-Klein eigenmodes.  In the next
sections we will study the conditions leading to the generation of
tachyon modes ($m^2<0$) that signal an instability of the model.

%%%%%%%%%%%%%%%%%%%%
\section{One brane}%
%%%%%%%%%%%%%%%%%%%%

It possible to construct models that include only one brane. 
In this case, the $Z_2$ symmetry emerges naturally, with all quantities
remaining regular in the second fixed point.
In this section, we will locate the brane at $y=\by$, and we
will assume the other fixed point $y_-$ to be at $y_-=0$. In this
case, the $Z_2$ symmetry at $y=0$ imposes
\begin{equation}\label{fiph0}
\vf_0'\left(0\right)=H\left(0\right)=0\,.
\end{equation}  
The constraint equation~(\ref{backcons}) then implies that the potential $V\left(\vf\right)$ has to vanish at $y=0$. Close to $y=0$, the solution of the Einstein equations~(\ref{backdyn}) reads
\begin{eqnarray}\label{backpert}
\vf_0\left(y\right)&=&\vf_0\left(0\right)+\frac{1}{2}\,a\left(0\right)^2\,
\frac{dV}{d\vf}(\vf_0(0))\,y^2+O\left(y^4\right)\,\,,\nonumber\\
H\left(y\right)&=&-\frac{\kappa^2}{9}\,a\left(0\right)^4\,
\left[\frac{dV}{d\vf}(\vf_0(0))\right]^2 y^3+O\left(y^5\right)\,\,.
\end{eqnarray}
We see from the last equation that $H\left(y\right)$ has to be
negative close to $y=0$. In fact, this turns out to be the case
everywhere:
\begin{equation}
\forall \,\, y \in \,\, ]\,0, y_+], \qquad H(y) <0\,.
\end{equation}
Indeed, if $H$ reaches zero at some $y=y_H$:
\vspace{0.1cm} \\
{\it(i)} either $\vf'_0\left(y_H\right)$ does not vanish, and then, by
eq.~(\ref{backdyn}),
$$
H'\left(y_H\right) <0 \,;
$$
{\it(ii)} or $\vf'_0\left(y_H\right)$ vanishes
(but then, $\vf''_0$ is non-zero, 
otherwise the bulk scalar field
would reduce to a cosmological constant). 
In this case, the derivatives
of eq.~(\ref{backdyn}) show that
$H'\left(y_H\right) = H''\left(y_H\right) = 0$, while
$$
H'''\left(y_H\right) = - \frac{2}{3} \, \kappa^2 \, {\vf_0''}^2\left(y_H\right) <0 \,.
$$
So, in both cases {\it (i)} and {\it (ii)}, 
$H$ should be turning from positive to negative, which
is in contradiction with the fact that $H\left(y<y_H\right)<0$. 

We will prove the existence of at least one tachyon mode in the
perturbations of this setup. To begin with, let us define a function
$F$ whose zeros will give the Kaluza-Klein mass spectrum:
\begin{equation}\label{defF}
F\left(m^2\right)\equiv g_{+}\,\Pi_{m^2}\left(\by\right)+m^2 \Psi_{m^2}\left(\by\right)\,\,,
\end{equation}
where
\begin{equation}\label{defPi}
\Pi_{m^2}\left(y\right) \equiv
\Psi_{m^2}'\left(y\right)+2\,H\left(y\right)\,\Psi_{m^2}\left(y\right)\,\,.
\end{equation}
In eqs.~(\ref{defF}, \ref{defPi}), $\Psi_{m^2}\left(y\right)$ is a solution of
eq.~(\ref{eqpert}) that satisfies the boundary
condition~(\ref{bounpert}) at $y=0$. Such boundary condition, as well
as the regularity of eq.~(\ref{eqpert}) in $y=0$, imposes that
$\Psi'_{m^2}\left(0\right)=0$ for every value of $m^2$ (this can also
be seen from eqs.~(\ref{const2},\ref{fiph0})). Without loss of
generality we set $\Psi_{m^2}\left(0\right)=1$~\footnote{Actually,
since the equation~(\ref{eqpert}) is singular for $y\rightarrow 0$,
there is an ambiguity in its solutions, and also a solution that goes
as $\Psi\sim y^3$ as $y\rightarrow 0$ would be acceptable. However,
since we require $\Psi\left(y\right)$ to be even under the exchange $y
\leftrightarrow -y$, the ambiguity is removed and we are left with a
solution that goes as $\Psi\sim 1 + m^2 y^2 / 2$.}. Also note that eq.~(\ref{eqpert}) can be conveniently re-expressed as a Hamilton--like system
\begin{eqnarray}\label{syst}
\Pi_{m^2}'
&=&
\left(-H +2\frac{\vf_0''}{\vf_0'}\right)\,\Pi_{m^2}
+\left(-m^2+\frac{2\kappa^2}{3}{\vf_0'}^2\right)\,
\Psi_{m^2}\,,\nonumber\\
\Psi_{m^2}'
&=&
\Pi_{m^2} -2\,H \,\Psi_{m^2}\,\,,
\end{eqnarray}
with the following initial conditions in $y=0$:
\begin{eqnarray}\label{ini_syst}
(\Psi_{m^2}, \Pi_{m^2})=(1,0), \qquad (\Psi'_{m^2}, \Pi'_{m^2})=(0,m^2) \,.
\end{eqnarray}

\vspace{0.2cm}

It turns out that $F(m^2)$ always goes through zero for some negative
value of $m^2$. To reach this result, one should first note that the
equation of motion for $\Psi_{m^2}$ can be explicitly
solved for $m^2\rightarrow -\infty$. In this limit, eq.~(\ref{eqv}) reduces to 
\begin{equation}
v_{m^2}''+m^2\,v_{m^2}-\frac{6}{y^2}\,v_{m^2}=0\,\,.
\end{equation}
We can solve this equation imposing the boundary condition
$\Psi\left(0\right)=1$. The solution reads
\begin{eqnarray}\label{solvinf}
v_{m^2}&=&\!\!\!-3\,
\frac{a(0)^{-1/2} m^2}{dV/d\vf(\vf_0(0))}\,\left[\left(1 - \frac{3}{m^2 y^2} \right) 
\cosh\left(|m|y\right)\, + \right.  
\nonumber\\
&&\qquad \left. - 3\,\frac{\sinh\left(|m|\,y\right)}{|m|\,y}\right]\cdot
\,\left(1+O\left(m^{-2}\right)\right)\,\,.
\end{eqnarray}
From this expression we can obtain the behavior of $\Psi_{m^2}$ for 
$|m|\, y \gg 1 $,
\begin{eqnarray}\label{psiminfdav}
\Psi_{m^2}(y)&=&-\frac{1}{2}\,\frac{a(0)^{3/2}}{a(y)^{3/2}}\,\frac{\vf_0'\left(y\right)}{a(0)^2\,dV/d\vf\,(\vf_0(0))}\,\e^{|m|\,y}\,\cdot\nonumber\\
&&\cdot\,\left[|m|+\left(\frac{\theta'\!\left(y\right)}{\theta\left(y\right)}-\frac{3}{y}\right)+O\left(|m|^{-1}\right)\right]\,\,, \nonumber\\
\Pi_{m^2}(y)&=& - 
\frac{|m|}{2} 
\,\frac{a(0)^{3/2}}{a(y)^{3/2}}\,\frac{\vf_0'\left(y\right)}{a(0)^2\,dV/d\vf\,(\vf_0(0))}\,\e^{|m|\,y}\,\cdot\nonumber\\
&&\cdot\,\left[|m| 
\, + \left( \frac{\kappa^2 {\vf_0'}^2}{3\,H} + \frac{3}{y} \right)
+ {\cal O}\left(|m|^{-1}\right)\right]\,\,.
\end{eqnarray}
In particular, for $y=y_+$, we have in the limit $|m|\, y_+ \gg 1 $
\begin{eqnarray}\label{psiinf}
&&\Psi_{m^2} \simeq -\frac{|m|}{2}\,\,\frac{a\left(0\right)^{3/2}}{a\left(y_+\right)^{3/2}} \, \, \frac{\vf_0'\left(y_+\right)}{a(0)^2\,dV/d\vf\,(\vf_0(0))}\,\,\e^{|m|\,y_+}\,\,,\nonumber\\
&&\Pi_{m^2}\simeq \,\,\,\,\,\,
|m| \, \Psi_{m^2}\,.
\end{eqnarray}
In the above equations, $dV/d\vf\,(\vf_0(0))$
has the same sign as $\vf_0'\left(y\right)$ for $y$ close to
$y=0$. If we assume that $\vf_0'$ vanishes in an odd
number of points in the range
$]\,0,\,y_+[$, this sign will be opposite
to the one of
$\vf_0'\left(y_+\right)$: so, 
$\Psi_{m^2}\left(y_+\right)$ 
and $\Pi_{m^2}\left(y_+\right)$ 
will be both positive. This implies that 
\begin{equation}
F(-\infty) = \lim_{m^2\rightarrow -\infty}\, 
m^2 \Psi_{m^2}\left(y_+\right) = - \infty \, .
\end{equation}
In the opposite case, i.e. when $\vf_0'$ vanishes in an even number
of points in the range $]\,0,\,y_+[$, we get that 
$F(-\infty) = + \infty$ (of course,
in the case ``even number of zeros'', we
include the configurations with no zeros at all).

\begin{figure}[ht]
\includegraphics[angle=0,width=0.45\textwidth]{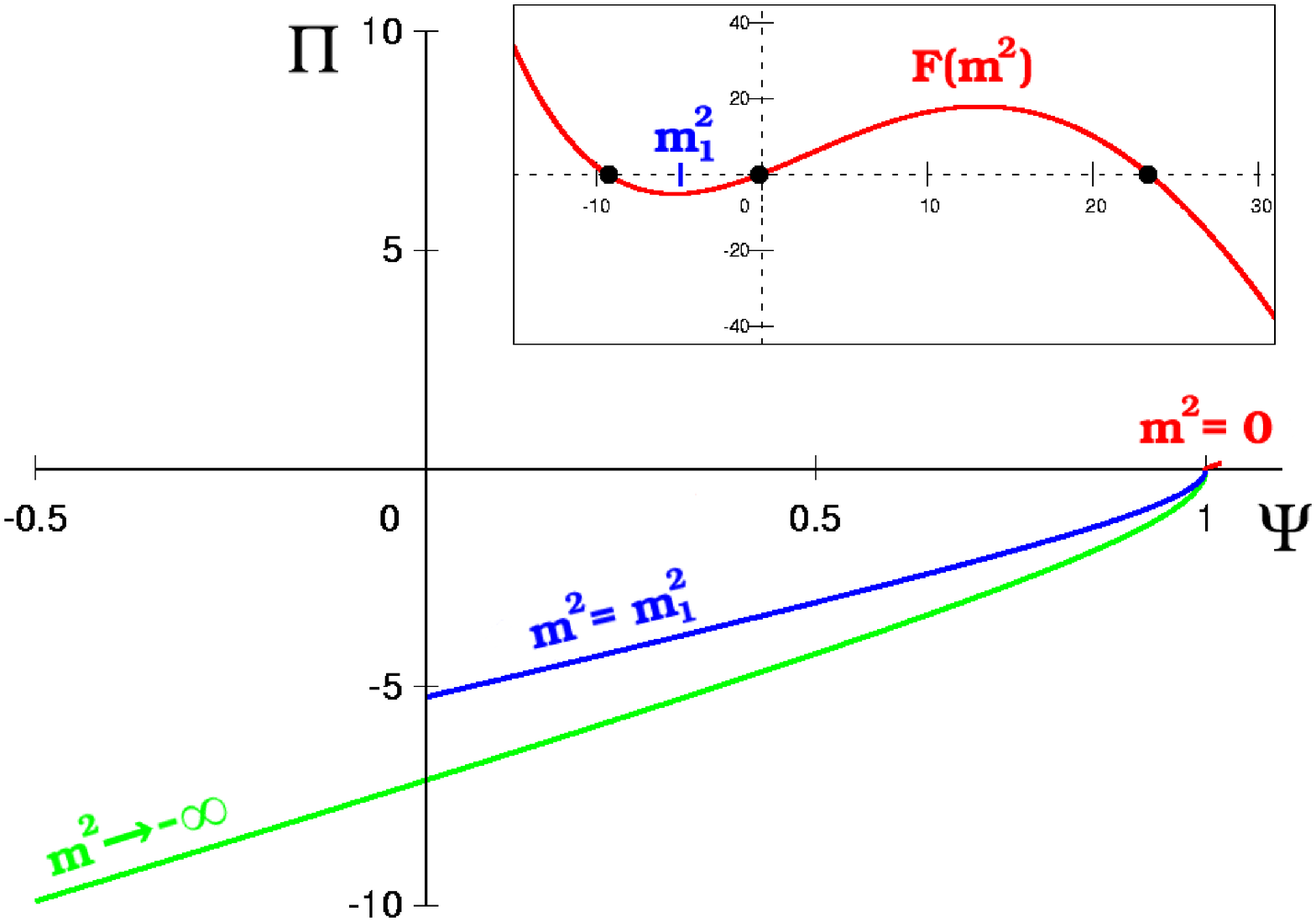}
\caption{\label{fig1} The phase--space trajectory for the three modes
$m^2=0,
m_1^2, -\infty$, for a particular single--brane background with
$V(\vf)= - V_0 + (m^2/2) \vf^2$, $U(\vf)= U_0 - (\mu^2/2) \vf^2$, and
$\vf_0(0) = (2 V_0 / m^2)^{1/2}$. In this example, there are no points
where $\vf_0'=0$. The zero--mode trajectory is the
very short red line starting from (1,0) and moving upward and right. 
The mode $m_1^2$ is defined in the text as the
first one hitting the axis $\Psi=0$, when $m^2$ decreases from 0 to
$-\infty$. On the upper right, we also show the function $F(m^2)$ for
the same model, in the case $g_+>0$. 
The three black dots correspond to $F(m^2)=0$, i.e.,
to the first three Kaluza-Klein modes. Two of them are tachyon modes.}
\end{figure}

The equation of motion~(\ref{eqpert}) or the system~(\ref{syst})
can be solved explicitly in a second
case, when $m^2=0$:
\begin{eqnarray}\label{psim0}
\Psi_0\left(y\right)&=&1-2\,\frac{H\left(y\right)}{a\left(y\right)^3}\,\int_0^y\,a\left(\tilde{y}\right)^3\,d\tilde{y}\,\,,\nonumber\\
\Pi_0\left(y\right)&=&\frac{2\,\kappa^2}{3}\,\frac{\vf_0'\left(y\right)^2}{a\left(y\right)^3}\,\int_0^y\,a\left(\tilde{y}\right)^3\,d\tilde{y}\,\,.
\end{eqnarray}  
This solution is such that $\Psi_0\left(y\right) \geq 1$, 
$\Psi_0'\left(y\right) \geq 0$ and $\Pi_0\left(y\right) \geq 0$.
So, $F(0) = g_+ \Pi_0(y_+)$ is of the same sign as $g_+$. 
On the phase--space diagram of
Fig.~\ref{fig1}, we plot the two solutions $m^2\rightarrow-\infty$
and $m^2=0$ 
obtained numerically for a particular choice of background.

\vspace{0.5cm}

For the last step of the proof, we should distinguish between three cases:
\vspace{0.2cm} \\
{\it(a)} if $\vf_0'$ has an even number of zeros in
the range $]\,0,\,y_+[$ and $g_+$ is negative, or $\vf_0'$ has an odd
number of zeros and $g_+$ is positive,
we immediately see that 
$F(0)$ and $F(-\infty)$ have opposite signs, so $F(m^2)$ goes
through zero for some negative value of 
$m^2$\footnote{Since the system is always well--behaved in terms of $v$,
$F$ is a continuous function of $m^2$.}, and the system has at least
one tachyon mode.
\vspace{0.2cm} \\ 
{\it(b)} if $\vf_0'$ has an even number of zeros in
the range $]\,0,\,y_+[$ and $g_+$ is positive, or
$\vf_0'$ has an odd number of zeros and $g_+$ is negative,
$F(0)$ and $F(-\infty)$ have the same sign, so we cannot 
draw an immediate conclusion. However,
it is not difficult to show that there exists at least
one value $m_1^2 < 0$ such that $F(m_1^2)$ and $F(0)$ are
of opposite sign.

The crucial point is that in the limit $m^2\rightarrow-\infty$, the
trajectory $(\Psi_{m^2}(y), \Pi_{m^2}(y))$ always includes some points
where $\Psi_{m^2}(y)$ is negative. When $\vf_0'$ has an even number of
zeros, this is obvious, since we know that $\Psi_{m^2}(y_+)$ must be
negative in the large $-m^2$ limit. When $\vf_0'$ has an odd number of
zeros, we will focus on the first of them, i.e., the smallest
coordinate value $y_0$ for which $\vf_0'(y_0)=0$.  The sign of
$\Psi_{m^2}(y_0)$ can be extracted in the limit of large $-m^2$ using
eq.~(\ref{psiminfdav}). The leading nonvanishing term comes from
$\theta'/\theta$, that diverges as $y\rightarrow y_0$, but is
multiplied by the overall factor $\vf_0'\left(y\right)$. We are thus
left with
\begin{equation}
\Psi_{m^2}\left(y_0\right)=\frac{a\left(0\right)^{3/2}}{a\left(y_0\right)^{3/2}}\frac{\vf_0''\left(y_0\right)}{a(0)^2 \, dV/d\vf(\vf_0(0))}\,\frac{\e^{|m|\,y_0}}{2}\,\,.
\end{equation}
In this expression, $dV/d\vf(\vf_0(0))$ has the same sign as
$\vf_0'\left(y\right)$ for $y$ close to zero, whereas
$\vf_0''\left(y_0\right)$ has the opposite sign. As a consequence,
$\Psi_{m^2}\left(y_0\right)$ turns out to be negative for large enough
$-m^2$.

This fact has important consequences.
When $m^2$ varies from zero to $-\infty$, the 
phase--space trajectory $(\Psi_{m^2}(y), \Pi_{m^2}(y))$ moves in a continuous
way from solution~(\ref{psim0}) to solution~(\ref{psiminfdav}). 
The former is such that $\Psi \geq 1$, while the latter includes points with
 $\Psi < 0$.
So, unavoidably,
there exists a value $m_1^2$ for which the trajectory just touches
the axis $\Psi=0$ for some value $\bar{y}$:
\begin{equation}\label{defm1}
\forall \, \, y, \quad \Psi_{m_1^2}(y) \geq 0 \qquad {\rm and} \qquad
\Psi_{m_1^2}(\bar{y})=0 \, .
\end{equation}
In $y=\bar{y}$, the derivative $\Psi_{m_1^2}'$ cannot be positive,
since the trajectory comes from the region $\Psi >0$.
So, $\Pi_{m_1^2}(\bar{y})=\Psi_{m_1^2}'(\bar{y})$ 
can be only negative or zero. Let us
discuss these two cases:

$\bullet$
if $\Pi_{m_1^2}(\bar{y})<0$, we find that $\bar{y}$ must be equal
to $y_+$, because otherwise $\Psi_{m_1^2}(y)$ would reach some negative
values for $y>\bar{y}$, in contradiction with the definition of $m_1^2$.

$\bullet$
if $\Pi_{m_1^2}(\bar{y})=0$, the trajectory reaches
the origin of phase--space. Let us suppose that $\bar{y}$ is a special
point where $\vf_0'$ vanishes. 
Then, the behavior around
the origin can be deduced directly from the perturbed Einstein equations
(\ref{const2}), (\ref{dyn}): it is easy to show that
$\Psi_{m_1^2}''(\bar{y})$ vanishes, but not $\Psi_{m_1^2}'''(\bar{y})$.
So,
\begin{equation}\label{origin}
\Psi_{m_1^2}(y) \simeq \alpha \, (y-\bar{y})^3 \,, \qquad  
\Pi_{m_1^2}(y) \simeq 3 \, \alpha \, (y-\bar{y})^2 \, ,
\end{equation}
where $\alpha$ is an unknown parameter.
This behavior cannot occur for 
the mode $m_1^2$, because $\Psi_{m_1^2}(y)$ would
change of sign in $y=\bar{y}$, in contradiction with the definition of 
$m_1^2$. So, $\bar{y}$ can only be in the range $]\,y_n, y_+]$
where $y_n$ is the last point in which
$\vf_0'$ vanishes. In this range, the trajectory
is governed by the system (\ref{syst}),
which tells that the trajectory will remain in $(\Psi_{m_1^2},
\Pi_{m_1^2}(y)) = (0,0)$ until $y=y_+$.

So, we have proved that in any case, the mode $m_1^2$
is such that 
\begin{equation} \label{defm12}
\Psi_{m_1^2}(y_+) = 0 \,, \qquad \Pi_{m_1^2}(y_+) \leq 0 \,.
\end{equation}
The trajectory corresponding to $m_1^2$ is shown in Fig.~\ref{fig1}
for a particular choice of background.  We finally obtain that
$F(m_1^2) = g_+ \Pi_{m_1^2}(y_+)$ is of sign opposite to $g_+$. Since
we know that $F(0)$ is of the sign of $g_+$, there is always at least
one tachyon mode with a squared mass in the range $m_1^2 \leq m^2 < 0$
(and in many cases, it is straightforward to prove the existence of a
second one in the range $- \infty < m^2 < m_1^2$).
\vspace{0.2cm} \\ 
{\it(c)} if $g_+=0$, or if $\vf_0'(y_+)=0$,
we see that $F(0)=0$: so, there
is a zero--mode, whose existence is phenomenologically unacceptable,
since it mediates (unobserved) long range forces, unless the $\Psi$
wave function is sufficiently suppressed on the brane. Precisely,
eq.~(\ref{psim0}) shows that $\Psi_0\left(y_+\right)$ must be bigger
than $\Psi_0\left(0\right)$ and than the wave function of the massless
graviton on the brane. As a consequence, $\Psi$ would carry long--range
forces stronger than gravity.

\vspace{0.5cm}

We insist on the fact that our conclusion doesn't rely on any particular
assumption concerning the background solution.
Therefore, any codimension--one compact model that includes one brane
at a fixed point of a $Z_2$ symmetry and a bulk scalar field will have
at least a tachyon mode, and will be unstable. This agrees with the
conjecture that was made in~\cite{TM} about the impossibility of
stabilizing models in which the background behaves at some point
as in eqs.~(\ref{backpert}) -- and proves it in the particular case of
compact models with only one brane.

%%%%%%%%%%%%%%%%%%%%%
\section{Two branes}%
%%%%%%%%%%%%%%%%%%%%%

In this section we will apply to systems of two branes techniques that are similar to the one used in the above section. We will deduce some properties of the potentials $V\left(\vf\right)$ and $U_\pm\left(\vf\right)$ that allow to establish whether any tachyon (or zero) modes exist the system.

To prove the absence of tachyon or zero modes, we simply require that the equations
\begin{equation}\label{bc1}
g_-\,\Pi_{m^2}\,\left(y_-\right)+m^2\,\Psi_{m^2}\left(y_-\right)=0
\end{equation}
and
\begin{equation}\label{bc2}
g_+\,\Pi_{m^2}\,\left(y_+\right)+m^2\,\Psi_{m^2}\left(y_+\right)=0
\end{equation}
cannot be solved simultaneously for $m^2\le 0$.

\subsection{Zero modes}

The existence of zero modes has been studied in~\cite{MK}. We review it here for completeness. First, it is immediate to see that if $g_+$ or $g_-$ vanish, then one of the equations~(\ref{bc1}) and~(\ref{bc2}) with $m^2=0$ is identically verified (whereas the other can be used to fix the boundary conditions), indicating the existence of a zero mode. Suppose now that $g_\pm\neq 0$. For $m^2=0$, eq.~(\ref{equ}) can be solved explicitly, yielding
\begin{equation}\label{u0}
u_0\left(y\right)=c_1\,\theta\left(y\right)+c_2\,\theta\left(y\right)\,\int_{y_-}^y\,\theta\left(\tilde{y}\right)^{-2}\,d\tilde{y}\,\,,
\end{equation}
that is,
\begin{eqnarray}\label{solm0}
\Psi_0\left(y\right)&=&\left[c_1-\frac{3}{\kappa^2}\,\frac{a\left(y_-\right)^3}{H\left(y_-\right)}\,c_2\right]\,\frac{H\left(y\right)}{a\left(y\right)^3}+\nonumber\\
&&+\frac{3}{\kappa^2}\,c_2\,\left[1-2\,\frac{H\left(y\right)}{a\left(y\right)^3}\,\int_{y_-}^y\,a\left(\tilde{y}\right)^3\,d\tilde{y}\right]\,\,\nonumber\\
\Pi_0\left(y\right)&=&-\frac{\kappa^2}{3}\,\frac{\vf_0'\left(y\right)^2}{a\left(y\right)^3}\,\left[c_1-\frac{3}{\kappa^2}\,\frac{a\left(y_-\right)^3}{H\left(y_-\right)}\,c_2\right]+\nonumber\\&&+2\,c_2\,\frac{\vf_0'\left(y\right)^2}{a\left(y\right)^3}\,\int_{y_-}^y\,a\left(\tilde{y}\right)^3\,d\tilde{y}\,\,.
\end{eqnarray}
Now, if $g_\pm\neq 0$, a zero mode exists if the equations $\Pi_0\left(y_-\right)=0$ and $\Pi_0\left(y_+\right)=0$ can be solved simultaneously with nonvanishing $c_1$ and $c_2$. The only case in which such condition is verified is if either $\vf_0'\left(y_-\right)$ or $\vf_0'\left(y_+\right)$ vanish. Therefore, to summarize, zero modes exist if and only if at least one of the conditions $g_+=0$, $g_-=0$, $\vf_0'\left(y_-\right)=0$ and $\vf_0'\left(y_+\right)=0$ holds. In no other case massless modes will emerge.

\subsection{A condition for stability}

Let us now turn to the existence of tachyon modes. We will assume that
the scale factor $a\left(y\right)$ is a monotonic function of $y$
(therefore the case in which $H=0$ at some point will not be considered in
the present and the following subsection), 
and we choose without loss of generality the
position of the branes in such a way that $H\left(y\right)<0$. We will
also assume that $\vf_0'$ does not vanish at any point (the case in
which $\vf_0'$ has zeros will be considered in 
the next subsection). Now
suppose that $g_->0$. Then eq.~(\ref{bc1}) implies that, for negative
$m^2$, $\Psi_{m^2}\left(y_-\right)$ and $\Pi_{m^2}\left(y_-\right)$
must have the same sign. Without loss of generality we can assume it
to be positive, and suppose $\Psi_{m^2}\left(y_-\right)=1$. Then the
second of eq.~(\ref{syst}) shows that, as long as
$\Pi_{m^2}\left(y\right)$ is positive, $\Psi_{m^2}\left(y\right)>1$.
But it is easy to see that $\Pi_{m^2}\left(y\right)$ can never get
negative. Indeed, to get negative, it should cross zero, but for
$\Pi_{m^2}=0$, the first of eqs.~(\ref{syst}) shows that (since
$\vf_0'\neq 0$ by hypothesis) at that point $d\Pi_{m^2}/dy>0$, so it
cannot be turning from positive to negative. Thus we conclude that for
whatever $m^2<0$, $\Psi_{m^2}\left(y_+\right)$ and
$\Pi_{m^2}\left(y_+\right)$ will be positive. As a consequence, if
$g_+$ is negative, eq.~(\ref{bc2}) cannot be fulfilled for any
negative $m^2$. Thus, for $g_->0$ and $g_+<0$, no tachyon modes can
exist and the system turns out to be {\it always} stable, as long as
the conditions $H\neq 0$ and $\vf_0'\neq 0$ hold. From the
definition~(\ref{defgpm}) we see that, if
$d^2U_\pm/d\vf_0^2\rightarrow +\infty$ the system is {\it always}
stabilized. In particular, this was the regime analyzed in~\cite{GW}.
Actually, from the above considerations it follows that whatever bulk
potential $V\left(\vf\right)$ 
which does not spoil the conditions $H\neq 0$ and $\vf_0'\neq 0$
can lead to radius stabilization if $d^2U_\pm/d\vf_0^2$ are positive and large enough.

\subsection{Instability of other cases}

Keeping the assumption that $a$ is a monotonic function of $y$, and
that the position of the branes is chosen in such way that $H(y)<0$,
we will show here that the three criteria formulated in the previous
subsection: $\vf_0'(y) \neq 0$, $g_->0$ and $g_+<0$, represent a
necessary and sufficient condition for stability.

Our strategy will be analogous to the one of the previous section, and
we will make use of the function $F\left(m^2\right)$ defined in
eq.~(\ref{defF}). To begin with, we observe that
$\Psi_{m^2}\left(y_-\right)$ cannot vanish. Indeed, if this was the
case the boundary condition~(\ref{bc1}) would impose that also
$\Psi'_{m^2}\left(y_-\right)$ vanishes (since we know that if
$g_\pm=0$ a zero mode exists, we will assume from now on that
$g_\pm\neq 0$). Then eq.~(\ref{eqpert}) would be trivially solved by
$\Psi=0$. We will thus be allowed to assume that the function
$\Psi_{m^2}$ appearing in the definition~(\ref{defF}) is a solution of
eq.~(\ref{eqpert}) such that $\Psi_{m^2}\left(y_-\right)=1$, and
$\Psi'_{m^2}\left(y_-\right)$ is determined by eq.~(\ref{bc1}).

We can evaluate $F\left(m^2\right)$ at $m^2=0$ and $m^2\rightarrow
-\infty$ as follows. For $m^2=0$, the solution of eq.~(\ref{eqpert})
is given in eq.~(\ref{solm0}). Imposing the boundary condition~(\ref{bc1}), we get
exactly the same solution $(\Psi_0(y), \Pi_0(y))$ as in eqs.(\ref{psim0}),
and we find that $F(0)$ is of the sign of $g_+$:
\begin{equation}\label{limm0}
F\left(0\right)=\left(\frac{2\,\kappa^2}{3}\,\frac{\vf_0'\left(y_+\right)^2}{a\left(y_+\right)^3}\,\int_{y_-}^{y_+}\,a\left(\tilde{y}\right)^3\,d\tilde{y}\right)\,g_+\,.
\end{equation}
Let us then turn to the evaluation of $F\left(m^2\rightarrow -\infty\right)$. 
In the limit of large $-m^2$, the 
equation for $v$ is trivial:
\begin{equation}\label{eqv2b}
v'' + m^2 v = 0 \,.
\end{equation}
We can then find the solution for $\Psi_{m^2}(y)$ using
eq.(\ref{vtopsi}), and normalize it to $\Psi_{m^2}(y_-)=1$ and to
eq.(\ref{bc1}). For $|m|\,(y-y_-) \gg 1$, the final solution reads
\begin{eqnarray}\label{psilargm}
&&\Psi_{m^2}(y) \simeq \frac{1}{g_-}\,\frac{a^{3/2}\left(y_-\right)}{a^{3/2}\left(y\right)}\,
\frac{\vf_0'\left(y\right)}{\vf_0'\left(y_-\right)}\, \left[|m| + \frac{ }{ }\right.
 \nonumber \\
&& \left.
+ \left(g_- + \frac{\theta'(y)}{\theta\,(y)} - \frac{\theta'(y_-)}{\theta\,(y_-)} \right)
+ {\cal O}(|m|^{-1})\right] \, 
\frac{\e^{|m|\,(y-y_-)}}{2}\,\,,\nonumber\\
&&\Pi_{m^2}(y) \simeq \frac{|m|}{g_-}\,\frac{a^{3/2}\left(y_-\right)}{a^{3/2}\left(y\right)}\,
\frac{\vf_0'\left(y\right)}{\vf_0'\left(y_-\right)}\, \left[|m| + \frac{ }{ }\right.
 \\
&& \left.
+ \left(g_- -\frac{\kappa^2 {\vf_0'}^2}{3\,H} - \frac{\theta'(y_-)}{\theta\,(y_-)}\right)
+ {\cal O}(|m|^{-1})\right] \, 
\frac{\e^{|m|\,(y-y_-)}}{2}\,\,, \nonumber
\end{eqnarray}
Inserting this expression in eq.~(\ref{defF}) gives, at leading order in $|m^2|$,
\begin{equation}\label{limminf}
\lim_{m^2\rightarrow -\infty}\!\!\!\!F\left(m^2\right)= 
m^2 \frac{|m|}{g_-}\,\frac{a^{3/2}\left(y_-\right)}{a^{3/2}\left(y_+\right)}\,
\frac{\vf_0'\left(y_+\right)}{\vf_0'\left(y_-\right)}\,
\frac{\e^{|m|\,(y_+-y_-)}}{2}\,\,,
\end{equation}
which is of opposite sign to $[g_- \vf_0'\left(y_+\right) / \vf_0'\left(y_-\right)]$.
We shall now distinguish between various cases:
\vspace{0.2cm} \\
{\it(a)} if $g_+$ and $[g_- \vf_0'\left(y_+\right) / \vf_0'\left(y_-\right)]$ have
the same sign, $F$ has a zero between $m^2=0$ and $m^2\rightarrow -\infty$, so there
is at least one tachyon mode.
\vspace{0.2cm} \\
{\it(b)} if  $g_+$ and $[g_- \vf_0'\left(y_+\right) / \vf_0'\left(y_-\right)]$ have
opposite signs, and $\vf_0'$ vanishes at least in one point $y_0$
in the range $]\,y_-,y_+[$,
we know that in the limit $m^2\rightarrow -\infty$, the solution for $\Psi_{m^2}(y)$ 
must change of sign in the vicinity of $y_0$, since $\vf_0'$ is in factor of 
eq.~(\ref{psilargm}).
So, we can use the same argument as in the single brane case: since $\Psi_{m^2}(y)$
is strictly positive for $m^2=0$ and goes to negative values for $m^2\rightarrow -\infty$,
there exists a mode $m_1^2$ obeying to the definition~(\ref{defm1}).
This mode 
must fulfill condition~({\ref{defm12}), and so
\begin{equation}
F(m_1^2) = g_+ \Pi_{m_1^2}(y_+)
\end{equation}
is of opposite sign to $g_+$ and to $F(0)$. So, $F$ has a root between
$m^2 = m_1^2$ and $m^2=0$, corresponding to a tachyon mode. 
\vspace{0.2cm} \\
{\it(c)} finally, if  $g_+$ and $[g_- \vf_0'\left(y_+\right) / \vf_0'\left(y_-\right)]$ have
opposite signs and $\vf_0'$ never vanishes -- which implies that 
$\vf_0'\left(y_+\right) / \vf_0'\left(y_-\right)$ is positive -- we can say that:
\vspace{0.2cm} \\
$\bullet$ either $g_+>0$ and $g_-<0$, and equation (\ref{psilargm}) shows that for
any value of $y$, $\Psi_{m^2}(y)$ is negative for  
sufficiently large $-m^2$. So, we can define the mode $m_1^2$ in the same way as in case
{\it (b)}, and reach the same conclusion that there is at least one tachyon mode. 
\vspace{0.2cm} \\
$\bullet$ or $g_+<0$ and $g_->0$, and we are exactly in the case of the previous subsection,
in which we know that there is no tachyon mode. 

\vspace{0.5cm}

The cases {\it (a)}, {\it (b)} and {\it (c)} include any possible situation. 
So, we conclude that
for two--brane systems with a monotonic scale factor $a(y)$, the necessary and sufficient
condition for radion stabilization reads
\begin{equation}\label{crit}
g_+ < 0, \quad g_- > 0, \quad  {\rm and} \quad \forall y, \quad \vf_0'(y) \neq 0\end{equation}
if the brane positions were chosen in such way that $H(y)<0$  -- the alternative choice $H(y)>0$ would lead to $g_+ > 0$, $g_- < 0$.

Criteria that guarantee radius stabilization were studied
in~\cite{KOP}, using an approach based on a four--dimensional
effective description of the system. The {\it extremization
  constraint} formulated in~\cite{KOP} can be rewritten in terms of
the quantities defined in the present paper as
\begin{equation}\label{kop}
\vf'_0\left(y_-\right)^2\,a\left(y_-\right)\,y_-^2\,g_--\vf'_0\left(y_+\right)^2\,a\left(y_+\right)\,y_+^2\,g_+\ge 0\,\,,
\end{equation}
where the inequality is saturated when there is a massless radion. This is indeed a necessary condition, since for $g_-$ positive and $g_+$ negative the inequality~(\ref{kop}) always holds. On the other hand, the criteria that lead to tachyon modes cannot be obtained by the above formula, that does not represent, as a consequence, a sufficient condition for radius stabilization. We attribute this to the fact that a four dimensional effective description is valid only if the system is actually close to a stable point.

Before giving an explicit example, we remark that the results discussed in the previous section for a single brane cannot be obtained as a special case of the present analysis. It is true that systems with a single brane can be seen as a special case of two brane systems, in two different ways. First, one can imagine that one of the two branes is fake and has a vanishing tension $U_-=0$. Second, one can impose $U_-=U_+$: then, there is a new symmetry with respect to $y = (y_+ + y_-)/2$, and the system is equivalent to a one--brane system plus its mirror image. In the first case, however, one would have $\varphi'\left(y_-\right)=0$ and therefore $g_-$ would be divergent. In the second case one would have that $a\left(y\right)$ is not monotonic between the two branes. In both cases, the analysis of the present section would not hold. This is why we have devoted a distinct section to the single brane case.

\subsection{An example}

As an application of our formalism, let us consider the Goldberger-Wise model~\cite{GW}. We consider a slight modification of the original model, replacing the quartic terms in the brane potentials by quadratic terms, such that the bulk and brane potentials read respectively
\begin{eqnarray}\label{exa}
V\left(\vf\right)&=&-\frac{6}{\kappa^2\,\ell^2}+\mu^2\,\vf^2\,\,,\nonumber\\
U_\pm\left(\vf\right)&=&\mp\frac{6}{\kappa^2\,\ell}+\frac{m_\pm}{2}\,\left(\phi-v_\pm\right)^2\,\,.
\end{eqnarray}
In the spirit of the original paper~\cite{GW}, we have assumed for the brane tensions their RS value $\mp 6/\left(\kappa^2\,\ell\right)$. However, consistency requires the sum $U_++U_-$ not to be positive~\cite{GKL}. Therefore, in general either $m_+$ or $m_-$ will have to be negative. Although negative mass terms are present in the five--dimensional lagrangian, this will not induce an instability as long as tachyonic modes do not appear in the four dimensional effective theory. Alternatively, we could have chosen to detune the brane tensions, thus allowing for a negative value of $U_++U_-$ even for positive values of both $m_+$ and $m_-$.
\begin{figure}[ht]
\includegraphics[angle=-90,width=0.45\textwidth]{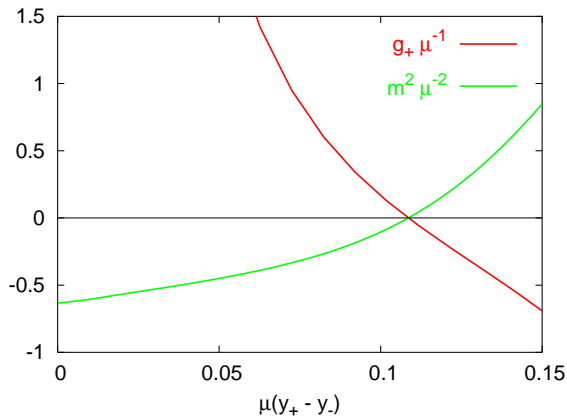}
\caption{\label{fig2} Value of $g_+$ and of the squared mass of the lightest Kaluza--Klein mode for the model~(\ref{exa}). The value of the various parameters can be found in the main text. The critical value $y_c$ below which the system is unstable is such that $y_c-y_-\simeq 0.11\,\mu^{-1}$.}
\end{figure}

We now fix $\ell$, $\mu$, $m_-$ and $v_-$, solve eqs.~(\ref{backisr}) in $y_-$ for $\vf$ and $\vf'$, and evolve the bulk equations~(\ref{backdyn}) until the position of the second brane $y_+$ is reached. There we read the value of $m_+$ and $v_+$. Therefore, we can deduce the behavior of $g_+$ as a function of $y_+$. In figure~\ref{fig2} such behavior is plotted for the following choice of parameters: $6/\left(\kappa^2\,\ell^2\right)=100$, $\kappa\,v_-=1$, $m_-=100\,\mu$. This implies that $g_-\simeq 35\,\mu$ is positive. The plot shows the presence of a critical value $y_c$ of $y_+$ such that, for $y_+<y_c$, $g_+$ has the same sign as $g_-$. Therefore, according the discussion of the preceding paragraphs, the system is unstable. On the other hand, for $y_+>y_c$, the system fulfills the stability criteria of the previous section, and no tachyon mode should appear.

As a cross check, in figure~\ref{fig2} we also plot the value of the squared mass of the first Kaluza--Klein mode (in units of $\mu^2$) as a function of $y_+$. Such plot is obtained by numerically searching the first zero of the function $F\left(m^2\right)$  defined in~(\ref{defF}). It is apparent that, when $g_+$ crosses zero, the squared mass of the first Kaluza--Klein mode turns from negative to positive. We thus see that the analytical criteria~(\ref{crit}) are verified by a full numerical computation. Moreover, for the model at hand, we see easily that there is a critical distance $y_c$ below which the model cannot be stable.

\vspace{0.5cm}

In conclusion, we have seen that the stability and instability of the
system we have considered is essentially encoded in the value of $g_+$
and $g_-$ (that depend only on background quantities). We have shown
that models including only one brane are unstable, independently of
the bulk and brane potentials. Zero modes emerge if and only if at
least one among the quantities $g_-$, $g_+$, $\vf_0'\left(y_-\right)$
and $\vf_0'\left(y_+\right)$ vanishes. 
Assuming that $H\left(y\right)$
is everywhere negative between $y_-$ and $y_+$, the condition for
stability is that $\vf_0'$ should never vanish
in this interval, while the quantities $g_-$ and $g_+$ defined
in eq.~(\ref{defgpm}) should be respectively positive and negative.

\section*{Acknowledgments}
The work of L.~S. is supported by the European
Community Human Potential program under contract HPRN-CT-2000-00152
``Supersymmetry and the Early Universe''. We would like to thank 
A.~Riotto for very interesting comments on this manuscript.


\begin{thebibliography}{99}

\bibitem{RS}
L.~Randall and R.~Sundrum,
%``A large mass hierarchy from a small extra dimension,''
Phys.\ Rev.\ Lett.\  {\bf 83} (1999) 3370
[arXiv:hep-ph/9905221].
%%CITATION = HEP-PH 9905221;%%

\bibitem{GW}
W.~D.~Goldberger and M.~B.~Wise,
%``Modulus stabilization with bulk fields,''
Phys.\ Rev.\ Lett.\  {\bf 83} (1999) 4922
[arXiv:hep-ph/9907447].
%%CITATION = HEP-PH 9907447;%%

\bibitem{TM}
T.~Tanaka and X.~Montes,
%``Gravity in the brane-world for two-branes model with stabilized  modulus,''
Nucl.\ Phys.\ B {\bf 582} (2000) 259
[arXiv:hep-th/0001092].
%%CITATION = HEP-TH 0001092;%%

\bibitem{CGK}
C.~Csaki, M.~L.~Graesser and G.~D.~Kribs,
%``Radion dynamics and electroweak physics,''
Phys.\ Rev.\ D {\bf 63} (2001) 065002
[arXiv:hep-th/0008151].
%%CITATION = HEP-TH 0008151;%%

\bibitem{pheno}
N.~Arkani-Hamed, S.~Dimopoulos and J.~March-Russell,
%``Stabilization of sub-millimeter dimensions: The new guise of the  hierarchy problem,''
Phys.\ Rev.\ D {\bf 63}, 064020 (2001)
[arXiv:hep-th/9809124];
%%CITATION = HEP-TH 9809124;%%
C.~Csaki, M.~Graesser, L.~Randall and J.~Terning,
%``Cosmology of brane models with radion stabilization,''
Phys.\ Rev.\ D {\bf 62}, 045015 (2000)
[arXiv:hep-ph/9911406];
%%CITATION = HEP-PH 9911406;%%
P.~Kanti, I.~I.~Kogan, K.~A.~Olive and M.~Pospelov,
%``Single-brane cosmological solutions with a stable compact extra  dimension,''
Phys.\ Rev.\ D {\bf 61}, 106004 (2000)
[arXiv:hep-ph/9912266];
%%CITATION = HEP-PH 9912266;%%
P.~Kanti, K.~A.~Olive and M.~Pospelov,
%``Static solutions for brane models with a bulk scalar field,''
Phys.\ Lett.\ B {\bf 481}, 386 (2000)
[arXiv:hep-ph/0002229];
%%CITATION = HEP-PH 0002229;%%
J.~Lesgourgues, S.~Pastor, M.~Peloso and L.~Sorbo,
%``Cosmology of the Randall-Sundrum model after dilaton stabilization,''
Phys.\ Lett.\ B {\bf 489}, 411 (2000)
[arXiv:hep-ph/0004086].
%%CITATION = HEP-PH 0004086;%%

\bibitem{trans}
G.~F.~Giudice, E.~W.~Kolb, J.~Lesgourgues and A.~Riotto,
%``Transdimensional physics and inflation,''
Phys.\ Rev.\ D {\bf 66}, 083512 (2002)
[arXiv:hep-ph/0207145].
%%CITATION = HEP-PH 0207145;%%

\bibitem{solu}
O.~DeWolfe, D.~Z.~Freedman, S.~S.~Gubser and A.~Karch,
%``Modeling the fifth dimension with scalars and gravity,''
Phys.\ Rev.\ D {\bf 62} (2000) 046008
[arXiv:hep-th/9909134].
%%CITATION = HEP-TH 9909134;%%

\bibitem{marco}
J.~Martin, G.~N.~Felder, A.~V.~Frolov, M.~Peloso and L.~Kofman,
%``Braneworld dynamics with the BraneCode,''
arXiv:hep-th/0309001.
%%CITATION = HEP-TH 0309001;%%

\bibitem{MK}
S.~Mukohyama and L.~Kofman,
%``Brane gravity at low energy,''
Phys.\ Rev.\ D {\bf 65} (2002) 124025
[arXiv:hep-th/0112115].
%%CITATION = HEP-TH 0112115;%%

\bibitem{KOP}
P.~Kanti, K.~A.~Olive and M.~Pospelov,
%``On the stabilization of the size of extra dimensions,''
Phys.\ Lett.\ B {\bf 538} (2002) 146
[arXiv:hep-ph/0204202].
%%CITATION = HEP-PH 0204202;%%


\bibitem{MFB}
V.~F.~Mukhanov, H.~A.~Feldman and R.~H.~Brandenberger,
%``Theory Of Cosmological Perturbations. Part 1. Classical Perturbations. Part 2. Quantum Theory Of Perturbations. Part 3. Extensions,''
Phys.\ Rept.\  {\bf 215} (1992) 203.
%%CITATION = PRPLC,215,203;%%

\bibitem{finelli}
F.~Finelli and R.~H.~Brandenberger,
%``Parametric amplification of gravitational fluctuations during  reheating,''
Phys.\ Rev.\ Lett.\  {\bf 82}, 1362 (1999)
[arXiv:hep-ph/9809490].
%%CITATION = HEP-PH 9809490;%%

\bibitem{mukhanov}
V.~F.~Mukhanov,
%``Quantum Theory Of Gauge Invariant Cosmological Perturbations,''
Sov.\ Phys.\ JETP {\bf 67}, 1297 (1988)
[Zh.\ Eksp.\ Teor.\ Fiz.\  {\bf 94N7}, 1 (1988)].
%%CITATION = SPHJA,67,1297;%%

\bibitem{GKL}
G.~W.~Gibbons, R.~Kallosh and A.~D.~Linde,
%``Brane world sum rules,''
JHEP {\bf 0101} (2001) 022
[arXiv:hep-th/0011225].
%%CITATION = HEP-TH 0011225;%%

\end{thebibliography}
\end{document}